\documentstyle[12pt,epsfig]{article}
\begin{document}
\baselineskip=20pt plus 1pt minus 1pt
\tolerance=1500
\vskip1cm
\begin{center}
{\large{\bf Alternative Linear Chiral Models for Nuclear Matter}} 
\end{center}
\vskip4mm
\begin{center}
A. Delfino$^1$, F. S. Navarra$^2$, M. Nielsen$^2$, R. B. Prandini$^2$,
M. Chiapparini$^3$
 \\
\vskip3mm
$^1$Instituto de F\'\i sica, Universidade Federal Fluminense,\\
Av. Litor\^anea s/n, 
24210-340, Boa Viagem, Niter\'oi, RJ, Brazil\\
\vskip3mm
$^2$Instituto de F\'\i sica da Universidade de S\~ao Paulo,\\
Caixa Postal 66318, 05315-970 S\~ao Paulo, SP, Brazil\\
\vskip3mm
$^3$Instituto de F\'\i sica, Universidade Estadual do Rio de Janeiro,\\
20559-900, Rio de Janeiro, RJ, Brazil\\
\vskip3mm
\end{center}

\vskip4mm

\noindent 
{\small{\bf Abstract:}  The equation of state of a family of
alternative linear chiral models in the mean field approximation is discussed. 
We investigate the analogy between some of these models with current models
in the literature, and we show that it is possible to reproduce very well
the saturation properties of nuclear matter.}

\vskip4mm


Since long time ago it has been known that chiral symmetry (CS) is
essential for the understanding of hadronic reactions. On the other
hand its role in nuclear physics was much less clear. The simplest
implementation of chiral symmetry in this context is the linear sigma
model \cite{gml}, where the scalar meson plays a dual role as the chiral
partner of the pion and the mediator of the intermediate-range nucleon-nucleon
attraction. However, this model was not able to reproduce the saturation
properties of nuclear matter at the mean field level and, therefore, was left 
in oblivion and much attention was given to field theoretical models like 
QHD \cite{sewa} 
in   its inumerous variants. During the eighties some important works tried
with relative success to cure the problems of the linear sigma model
through the introduction of a vector-scalar coupling \cite{bog}. However,
these improved versions of the model were shown \cite{fur} to fail in
giving a good description of finite nuclei properties. The hope, and to some
extent increasing evidence \cite{fri}, that chiral symmetry is really 
important for nuclear physics motivated the implementation of nonlinear 
realizations of  CS, some of them being very
successfull in the applications to finite nuclei \cite{tan,fo}. 

Alternatively, with  the introduction of a scalar glueball field, 
a reasonable description of nuclear matter and
closed shell nuclei was achieved in the context of the linear
sigma model \cite{hrl}. The idea behind the introduction of the glueball
field was to create an effective lagrangian that respects the low-energy
theorems of scale and chiral symmetry. The authors of ref.\cite{hrl} have
also noticed that the coupling of the vector field with the glueball
field rather than the sigma field is essential to provide a good description
of finite nuclei, in agreement with the conclusions in \cite{fur}.

Prospectively,
in this paper we show that it is possible to get a good description of
nuclear matter and eventually of finite nuclei with a generalized 
formulation of the
linear sigma model, that contains only the original scalar, pseudoscalar
and vector meson fields without additional couplings between the mesons.
To generate the model we use, as motivation, the quark meson coupling 
model (QMC)
\cite{st} which is a model for nuclear matter and finite nuclei in which
the quark structure of the nucleons is explicitly considered. The basic
assumptions of the model are that effective meson fields couple directly
to quarks confined in non-overlaping bags, and that nucleons obey the Dirac 
equation in the effective meson mean fields whose sources are the quarks in 
the bag. The QMC model shares important features with the Walecka model (QHD),
such as the relativistic saturation mechanism. The final equation for the total
energy per nucleon in nuclear matter is identical to that of QHD. The effect
of internal quark structure enters only through the effective mass and
the self-consistency condition for the $\sigma$ field. The self-consistency 
condition,  on the other hand, differs from that of QHD through a scalar
density factor. This is equivalent of having a coupling constant which 
depends on the scalar field and consequently on the density. A similar 
dependence is also found in the chiral confining model of ref. \cite{ban}. 
In that work it is also shown via a Dirac-Brueckner-Hartree-Fock 
calculation that  
significant changes in the nuclear matter saturation may be observed  
when the mesonic coupling constants  depend on the density. Another kind of
density dependent coupling constant was demonstrated to come from
relativistic $SU(6)$ model \cite{kiro}.

Since it is well known that models with linear realization of chiral 
symmetry have 
difficulty to saturate the nuclear matter we pose the question whether a 
density dependence of the mesonic coupling constants, claimed in the above 
mentioned works, could   
provide a reasonable saturation mechanism in this class of chiral models.
Therefore, we will allow the coupling constants in the linear 
sigma model to be dependent of the scalar field. We start with the 
following lagrangian: 
\begin{eqnarray}
{\cal L}\,&=&\,\overline\Psi[\gamma_\mu(i\partial^\mu-g_v^* V^\mu) - 
g_\sigma^*(\sigma+i\gamma_{5}{\vec\tau}\cdot{\vec\pi})]\Psi\,
+\,\frac{1}{2}(\partial_\mu\sigma\partial^\mu\sigma - 
\partial_\mu{\vec\pi}\cdot\partial^\mu{\vec\pi}) 
\nonumber \\[0.3cm]  
 & &-\frac{1}{4}F_{\mu\nu}F^{\mu\nu} + \frac{1}{2}m_v^2 V_\mu V^\mu -
\frac{1}{4}\lambda(\sigma^2 + \pi^2 -u^2)^2
\;  ,  
\label{la}
\end{eqnarray}
where $F_{\mu\nu}=\partial_\mu V_\nu-\partial_\nu V_\mu$. Here
$\Psi$, $V_\mu$, $\sigma$ and $\pi$ are respectively the nucleon,
vector meson, scalar meson and pion field. $ u $ is the vacuum
expectation value of the sigma field which gives mass to the nucleon.
In this version of the sigma model there is no symmetry breaking term
and the symmetry is realized in the Nambu-Goldstone mode. The effective
coupling constants are given by
\begin{eqnarray}
g_\sigma^*&=& {g_\pi\over \left(1+{\sigma^2 + \pi^2 -u^2\over u^2}
\right)^\alpha}
\; ,\label{gs1}
\\
g_v^*&=& {g_v\over \left(1+{\sigma^2 + \pi^2 -u^2\over u^2}\right)^\beta}
\; ,
\label{gv1}
\end{eqnarray}
where $g_\pi$ is the pion-nucleon coupling contant, which value is
fixed at the experimental value: ${g_\pi^2\over4\pi}=13.6$, and $g_v$, 
the vector meson-nucleon coupling constant, is a free parameter. $\alpha$
and $\beta$ are constants which define different families of models. After 
the introduction of field dependent couplings the final functional dependence 
of the lagrangian density on the fields is very different from the original 
linear $\sigma$ model. However we preserve the linear realization of the 
chiral symmetry. This is explicit in the new factors which depend only on the 
combination $\sigma^2 + \pi^2$.

We now shift
the scalar field defining $\sigma=u+s$ , such that 
$\langle s \rangle=0$ in the vaccum. The lagrangian becomes:
\begin{eqnarray}
{\cal L}\,&=&\,\overline\Psi[\gamma_\mu(i\partial^\mu-g_v^* V^\mu) -M^*
-i g_\sigma^*\gamma_{5}{\vec\tau}\cdot{\vec\pi}]\Psi\,
+\,\frac{1}{2}(\partial_\mu s\partial^\mu s - m^2_s s^2)
\nonumber \\[0.3cm]
 &+ &\frac{1}{2}\partial_\mu{\vec\pi}\cdot\partial^\mu{\vec\pi}   
-\frac{1}{2}F_{\mu\nu}F^{\mu\nu} + \frac{1}{2}m_v^2 V_\mu V^\mu -
{g_\pi m_s^2\over2M} s (s^2+\pi^2) -{g_\pi^2m_s^2\over8M^2}
(s^2+\pi^2)^2 \;  , 
\end{eqnarray}
where $m^2_s=2\lambda u^2$, $M=g_\pi u$
and $M^*={g_\sigma^*\over g_\pi}(M+g_\pi s)$. The non vanishing vaccum 
expectation value
of the $\sigma$ field generates both the $\sigma$ and nucleon masses. 
In a mean field (MF) approximation the energy density is given by
\begin{eqnarray}
{\cal E}_{MF}\,=\,\frac \gamma{\left( 2\pi \right)^3}
\int\limits_0^{k_F}d^3k \sqrt{k^{2}+M^{*2}}+ \frac{C_{v}^2}{2M^2}
\rho_B^2 \left({M^*\over M}\right)^{4\beta\over2\alpha-1}
+\frac{M^4}{8C_\sigma^2}\left[\left({M^*\over M}\right)^{-2\over2\alpha-1}
-1\right]^2\; ,
\label{mf}
\end{eqnarray}
where $C_\sigma^2={g_\pi^2M^2\over m_s^2}$, $C_v^2={g_v^2M^2\over 
m_v^2}$,  $\rho _B$ is the barion density:
\begin{equation}
\rho _B=\frac \gamma{\left( 2\pi \right)^3}\int\limits_0^{k_F}d^3k\; ,
\end{equation}
and $\gamma$ is the spin-isospin degeneracy ($\gamma=4$ for nuclear matter).

To write Eq.(\ref{mf}) we have used the fact that in MF we can write the
effective coupling constants as
\begin{eqnarray}
g_\sigma^*&=& g_\pi\left({M^*\over M}\right)^{2\alpha\over2\alpha-1}
\; ,\label{gs}\\
g_v^*&=& g_v\left({M^*\over M}\right)^{2\beta\over2\alpha-1}
\; .\label{gv}
\end{eqnarray}

The effective nucleon mass, $M^*$, is determined by minimizing Eq.~(\ref{mf})
with respect to $M^*$:
\begin{eqnarray}
2\beta C_v^2{\rho_B^2\over M^6}\left({M^*\over M}\right)^{4\beta+2-4\alpha
\over2\alpha-1}&-&{1\over2C_\sigma^2}\left({M^*\over M}\right)^
{-4\alpha\over2\alpha-1}\left(\left({M^*\over M}\right)^{-2\over
2\alpha-1}-1\right)
\nonumber\\
&=&(1-2\alpha)\frac \gamma{M^2}\int\limits_0^{k_F}
{d^3k\over\left( 2\pi \right)^3} {1\over\sqrt{k^2+M^{*2}}}\; .
\label{gap}
\end{eqnarray}         

At this point it is interesting to compare the MF result of our model with 
models existing in the literature. In particular, in ref.\cite{mos1} a
unified model Lagrangian density was proposed to connect QHD with the 
Zimanyi-Moszkowski (ZM) model \cite{zm} and a modified version of the ZM model,
called ZM3 in ref.\cite{mos2}. In the MF approach, the energy
density of this unified model is given by
\begin{eqnarray}
{\cal E}_{un}\,=\,\frac \gamma{\left( 2\pi \right)^3}
\int\limits_0^{k_F}d^3k \sqrt{k^{2}+M^{*2}}+ \frac{C_{v}^2}{2M^2}
\rho_B^2 \left({M^*\over M}\right)^a
+\frac{M^4}{2C_\sigma^2}\left({1-M^*/M\over (M^*/M)^b}\right)^2 \; ,
\label{mfzm}
\end{eqnarray}
where the gap equation for the effective nucleon mass is  
\begin{eqnarray}
{a C_v^2\over2}{\rho_B^2\over M^6}\left({M^*\over M}\right)^{a-b-1}
-{1\over C_\sigma^2}\left({M^*\over M}\right)^
{-3b-1}\left(1-{M^*\over M}\right)
=-\frac \gamma{M^2}\int\limits_0^{k_F}
{d^3k\over\left( 2\pi \right)^3}  {1\over\sqrt{k^2+M^{*2}}} \; .
\label{gapzm}
\end{eqnarray}         

The values of $a$ and $b$ for the different models given in Eq.(\ref{mfzm})
are given by
\begin{center}
\begin{tabular}{|c|c|c|}  \hline
model & a & b \\
\hline
QHD & 0 & 0 \\
\hline
ZM & 0 & 1 \\
\hline
ZM3 & 2 & 1 \\
\hline
\end{tabular}
\end{center}

Comparing Eqs.(\ref{mf}) and (\ref{mfzm}) and Eqs.(\ref{gap}) and (\ref{gapzm})
we can see that in MF our model can reproduce QHD, ZM and ZM3 models if 
$\alpha$ and $\beta$ assume the values given bellow, provided that we 
redefine our $C_\sigma^2$ as $C_\sigma^2={g_\pi^2M^2\over m_s^2/4}$. 
\begin{center}
\begin{tabular}{|c|c|c|}  \hline
model & $\alpha$ & $\beta$ \\
\hline
QHD & -1/2 & 0 \\
\hline
ZM & 3/2 & 0 \\
\hline
ZM3 & 3/2 & 1 \\
\hline
linear-$\sigma$ model & 0 & 0 \\
\hline
\end{tabular}
\end{center}
We have included in the table above the values of $\alpha$ and $\beta$
that reproduce the original linear-$\sigma$ model \cite{gml,sewa,ben}.
As it is well known, the original linear-$\sigma$ model
does not reproduce the nuclear matter ground-state properties. This happens
because in this model $M^*$ goes to zero at small values of the nuclear
density and this generates a very strong attraction. In our model this 
excess of attraction is avoided by the decrease of the scalar coupling
constant as a function of the density, as can be seen in Eq.(\ref{gs})
(since $M^*$ is a decreasing function of the density) for $\alpha<0$ and 
$\alpha>1/2$. 
\begin{figure}[h]
\begin{center}
\vskip -1.5cm
\centerline{\epsfig{figure=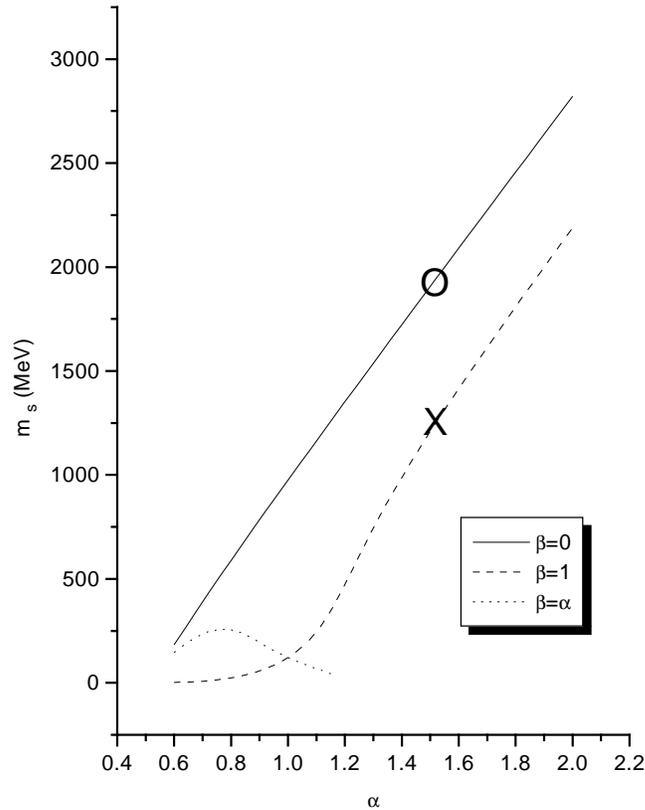,width=10cm}}
\vskip -1.cm
\caption{\footnotesize The scalar meson mass as a function of $\alpha$ for
different values of $\beta$. The circle and the cross stand for the
ZM and ZM3 results respectively.}
\end{center}
\label{fig1}
\end{figure}

We will analyze our model in the region $\alpha>1/2$ and for three choices
of $\beta$: $\beta=0$, $\beta=1$ and $\beta=\alpha$. The values of $C_v^2$
and $C_\sigma^2$ are fixed to reproduce the energy per nucleon
$E/N=-15.75$ MeV at $k_F=1.3$ fm$^{-1}$. Since $g_\pi$ is fixed at
its experimental value, different values of $C_\sigma^2$ are related
with different values of $m_s$. In Fig.1 we show the value of the scalar 
meson mass obtained from $C_\sigma^2$ that saturates the nuclear matter,
as a function of $\alpha$. The particular results obtained in the ZM and 
ZM3 models are assigned in this figure by a circle and a cross respectively.
From this figure we can see that the choice $\beta=\alpha$ provides very
smaller values for $m_s$ and delimits the region where the symmetric
scaling ($\alpha=\beta$) is valid. If one thinks about finite systems,
$m_s$ is associated with the range of the scalar interaction. Therefore,
smaller values of  $m_s$ provide longer  ranges of the interaction
in the configuration space.
\begin{figure}[h]
\begin{center}
\vskip -1.5cm
\centerline{\epsfig{figure=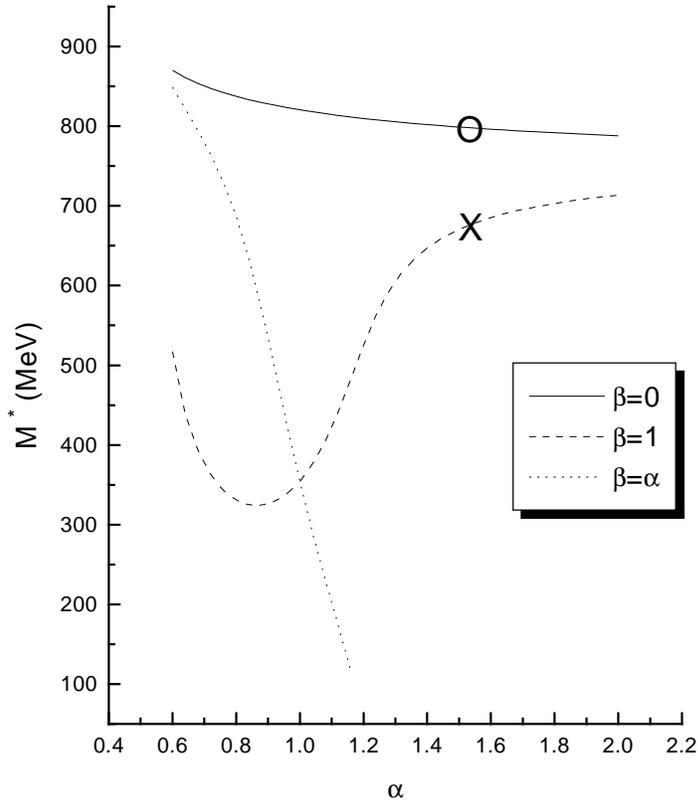,width=10cm}}
\vskip -1.cm
\caption{\footnotesize The effective nucleon mass at saturation density
as a function of $\alpha$ for different values of $\beta$. The circle and the 
cross stand 
for the ZM and ZM3 results respectively.}
\end{center}
\label{fig2}
\end{figure}
\begin{figure}[h]
\begin{center}
\vskip -1.5cm
\centerline{\epsfig{figure=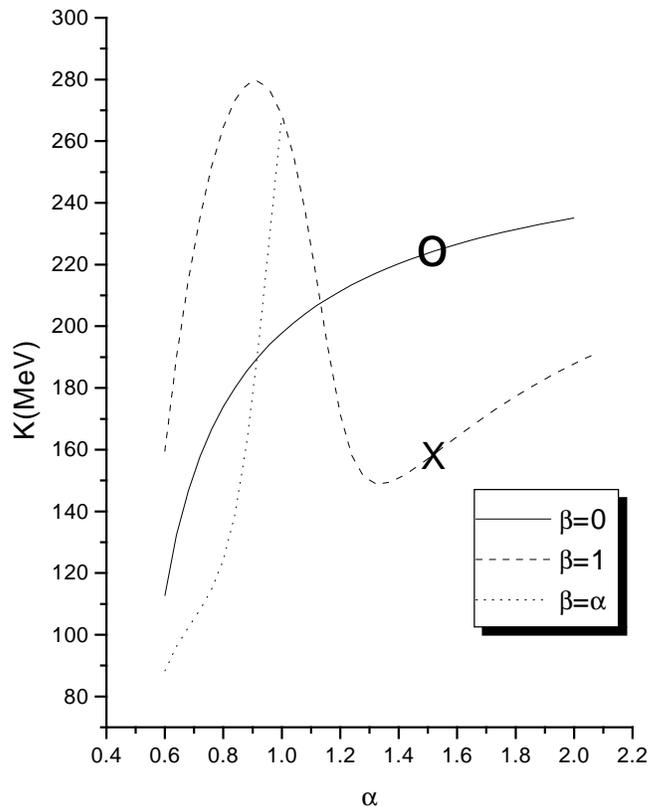,width=10cm}}
\vskip -1.cm
\caption{\footnotesize The incompressibility as a function of 
$\alpha$ for different values of $\beta$. The circle and the cross stand 
for the ZM and ZM3 results respectively.}
\end{center}
\label{fig3}
\end{figure}

In Figs.2 and 3 we show the effective nucleon mass at saturation density, 
$M_0^*$, and the incompressibility, $K$, as a function of $\alpha$. 
As we can see by these figures, a large region of values for 
$K$ and $M_0^*$ can be generated by these models and we can choose
the values of  $\alpha$ and $\beta$ that can better reproduce
the phenomenology. It is also interesting to notice that the values
generated for the incompressibility by these models are generally small,
as compared with QHD for instance.

There is a strong correlation in relativistic models between the effective 
nucleon  mass at saturation density, $M_0^*$, and the strenght of the
spin orbit force in nuclei \cite{rein,bod,frs}. Typical values of $M_0^*/M$ 
in successful models are roughly 0.6, which yield spin-orbit splittings close 
to experimental values. Accurate reproductions of observables in finite nuclei,
including the spin-orbit splitting in particular, tightly constrain the value
of $M_0^*/M$ in the range $0.58\leq M_0^*/M\leq 0.70$ \cite{frs}. Since in 
linear chiral models in general $M_0^*/M$ is too close to unity, this was
found as one of the reasons for the failure of the models considered in
ref.\cite{fur} to give a good description of finite nuclei. As can be seen by 
figure 2, allowing the coupling constants to decrease with the density
we can generate a linear chiral model which gives $M_0^*/M$ in good
agreement with phenomenology. We can also see that the choice $\beta=0$
does not go in this direction generating $M_0^*/M$ very close to unity,
indicating its almost non-relativistic character.
The choice $\alpha=\beta$ and $\beta=1$ (for $\alpha$ around 1.5 (ZM3))
seems to be a good candidate.
\begin{figure}[h]
\begin{center}
\vskip -1.5cm
\centerline{\epsfig{figure=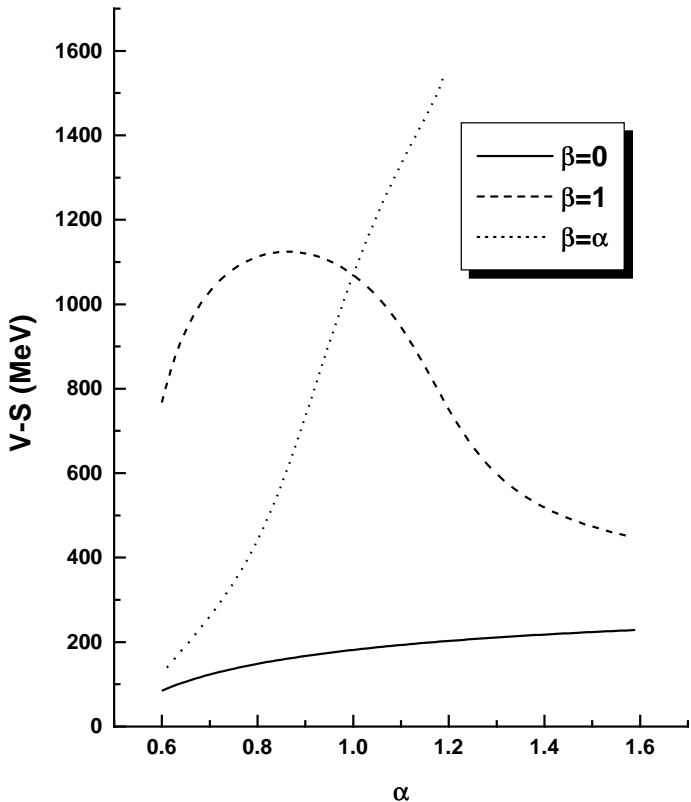,width=10cm}}
\vskip -1.cm
\caption{\footnotesize $V-S$ as a function of 
$\alpha$ for different values of $\beta$. }
\end{center}
\label{fig4}
\end{figure}

Another qualitative ingredient associated with the spin-orbit splitting 
in finite nuclei
is the difference between the depth of the vector and scalar  potencials,
$V-S$ ($S=M^*-M$, $V={g_v^*}^2\rho_B/m_v^2$). 
The higher the difference the higher the splitting. In Fig. 4 we present
the difference $V-S$ as a function of $\alpha$. Typical values of this
quantity, that give reasonable values for the spin-orbit splitting,
lies usually in the region from 550 to 650 MeV \cite{tan}. In our models
this region can be obtained for the cases $\alpha=\beta$ and $\beta=1$
in agreement with the analysis done above. 
Of course this is only  an indication that good results can be obtained
for finite nuclei. A calculation for finite nuclei with these models is under 
way \cite{novo}.

To summarize, we have proposed a new class of models for nuclear systems
where chiral symmetry is implemented in a linear realization. No new
degress of freedom or new interactions other than that present in the original
linear-$\sigma$ model \cite{gml,mase} were needed. In these models loops
are not an essential ingredient to obtain saturation of nuclear matter
as in refs.\cite{mase,ben}. We found that this class of models
reproduces very well the expected saturation properties of nuclear matter.
In a mean field approximation and for some particular values of the 
parameters, these models can reproduce the
well known derivative coupling ZM models and QHD. Therefore ZM models 
and QHD can also be  understood as reminiscents of 
linear chiral models  with couplings given by Eqs. (\ref{gs1}) and 
(\ref{gv1}), after a mean field approximation and proper choice of 
$\alpha$ and $\beta$. Let's remark that ZM models have a desirable
nonrelativistic limit in the saturation mechanism. We are aware that for 
finite nuclei, surface terms
can not be neglected and the results for nuclei spectra may differ from
the original models \cite{mos1}. Finally we would like to point out that to 
the best of our knowledg it is the first time that  this class of linear
chiral models can reproduce well nuclear matter phenomenology.

\vskip2mm

\noindent
{\bf ACKNOWLEDGEMENTS}
\vskip1mm

This work was partially supported by FAPESP and CNPq -- Brazil.

\eject



\begin{thebibliography}{99}

\bibitem{gml} M. Gell-Mann and M. L\'evy, {\sl Nuovo Cimento} {\bf 16}, 705 
(1960).

\bibitem{sewa} B.D. Serot and J. D. Walecka, {\sl  Adv. Nucl. Phys.} {\bf
16}, 1 (1986); B.D. Serot, Rep. Prog. Phys. {\bf55}, 1855 (1992).

\bibitem{bog} J. Boguta, {\sl Phys. Lett.} {\bf B120}, 34 (1983);
{\bf B128}, 19 (1983).

\bibitem{fur} R.J. Furnstahl and B.D. Serot,  {\sl Phys. Rev.} {\bf C47},
2338 (1993);{\sl Phys. Lett.} {\bf B316}, 12 (1993).

\bibitem{fri} J. L. Friar, Few-Boby Systems Suppl. {\bf 99} (1996) 1 and
references therein.

\bibitem{tan} R.J. Furnstahl, H.B. Tang and B.D. Serot,  
{\sl Phys. Rev.} {\bf C52}, 1368 (1995); R.J. Furnstahl, B.D. Serot and
H.B. Tang, {\sl Nucl. Phys.} {\bf A615}, 441 (1997).

\bibitem{fo} V.N. Fomenko et al., {\sl J. Phys.} {\bf G 21 }, 53 (1995).

\bibitem{hrl}  E.K. Heide, S. Rudaz and P.J. Ellis,
{\sl Nucl. Phys.} {\bf A571}, 713 (1994); P.J. Ellis, E.K. Heide and S. 
Rudaz, {\sl Phys. Lett.} {\bf B282}, 271 (1992).

\bibitem{st} K. Saito and A.W. Thomas, {\sl Phys. Lett.} {\bf B327}, 9 (1994);
K. Saito, K. Tsushima and A.W. Thomas, {\sl Nucl. Phys.} {\bf A609}, 339
 (1996).

\bibitem{ban} M.K. Banerjee and J.A. Tjon, 
{\sl Phys. Rev.} {\bf C56}, 497 (1997).

\bibitem{kiro} K. Miyazaki,  {\sl Prog. Theor. Phys.}  {\bf 91}  1271 (1994).


\bibitem{mos1} A. Delfino, M. Chiapparini, M. Malheiro, L.V. Belvedere
and A.O. Gattone, {\sl Z. Phys.} {\bf A355}, 145 (1996); M. Chiapparini, 
A. Delfino, M. Malheiro and A.O. Gattone, {\sl Z. Phys.} {\bf A357}, 47 (1997).

\bibitem{zm} J. Zimanyi and S.A. Moszkowski, {\sl Phys. Rev.} {\bf C42},
1416 (1990).

\bibitem{mos2} A. Delfino, C.T. Coelho and M. Malheiro, {\sl Phys. Lett.} 
{\bf B345}, 361 (1995); {\sl Phys. Rev.} {\bf C51}, 2188 (1995);

\bibitem{ben} W. Bentz, L.G. Liu and A. Arima,  {\sl Ann. Phys. (N.Y.)} 
{\bf 188}, 61 (1988) ; L.G. Liu, W. Bentz and A. Arima, 
{\sl Ann. Phys. (N.Y.)}{\bf 194}, 387 (1989).

\bibitem{rein} P.-G. Reinhard, {\sl Rep. Prog. Phys.} {\bf52}, 439 (1989).

\bibitem{bod} A.R. Bodmer, {\sl Nucl. Phys.} {\bf A526}, 703 (1991).

\bibitem{frs} R.J. Furnstahl, J.J. Rusnak and B.D. Serot, {\sl Nucl. Phys.} 
{\bf A632}, 607 (1998).

\bibitem{caco} C. Callan, S. Coleman, J. Wess and B. Zumino, {\sl Phys. Rev.} 
{\bf 177}, 2247 (1969).

\bibitem{novo} M. Chiapparini, A. Delfino, F. S. Navarra, M. Nielsen, work in
progress.

\bibitem{mase} T. Matsui and B.D. Serot, {\sl Ann. Phys.} {\bf 144}, 107 
(1982).





\end{thebibliography}
\end{document}